\documentclass[structabstract]{aa}

\usepackage{graphicx}
\usepackage{txfonts}
\usepackage[round]{natbib}
\bibpunct{(}{)}{;}{a}{}{,} 
\usepackage{epsfig}

\begin{document}

   \title{The dusty environment of HD 97300\\as seen by
     \emph{Herschel} and \emph{Spitzer}\thanks{This work is based on
       observations made with the {\it Herschel} Space Observatory and
       with the \emph{Spitzer} Space Telescope. \emph{Herschel} is an
       ESA space observatory with science instruments provided by
       European-led Principal Investigator consortia and with
       important participation from NASA. \emph{Spitzer} is operated
       by the Jet Propulsion Laboratory, California Institute of
       Technology under a contract with NASA.}}

   \author{\'A. K\'osp\'al
          \inst{1}
          \and
          T. Prusti\inst{1}
          \and
          N. L. J. Cox\inst{2}
          \and
          G. L. Pilbratt\inst{1}
          \and
          Ph. Andr\'e\inst{3}
          \and
          C. Alves de Oliveira\inst{4}
          \and
          E. Winston\inst{1}
          \and 
          B. Mer{\'\i{}}n\inst{4}
          \and
          A. Ribas\inst{4}
          \and
          P. Royer\inst{2}
          \and
          R. Vavrek\inst{4}
          \and
          C. Waelkens\inst{2}
          }

   \institute{Research and Scientific Support Department, European
     Space Agency (ESA-ESTEC, SRE-SA), PO Box 299, 2200 AG, Noordwijk,
     The Netherlands\\
     \email{akospal@rssd.esa.int}
     \and
     Instituut voor Sterrenkunde, KU Leuven, Celestijnenlaan 200D,
     B-3001, Leuven, Belgium
     \and
     Laboratoire AIM Paris--Saclay, CEA/DSM--CNRS--Universit\'e Paris
     Diderot, IRFU, Service d'Astrophysique, Centre d'Etudes de
     Saclay, Orme des Merisiers, 91191 Gif-sur-Yvette, France
     \and
     Herschel Science Centre, ESA-ESAC, PO Box 78, E-28691 Villanueva
     de la Ca\~nada, Madrid, Spain}

   \date{Received date; accepted date}

   \abstract
   {}
   {We analyze the surroundings of HD\,97300, one of two
     intermediate-mass stars in the Chamaeleon\,I star-forming
     region. The star is known to be surrounded by a conspicuous ring
     of polycyclic aromatic hydrocarbons (PAHs).}
   {We present infrared images taken with \emph{Herschel} and
     \emph{Spitzer} using 11 different broad-band filters between
     3.6$\,\mu$m and 500$\,\mu$m. We compare the morphology of the
     emission using cuts along different position angles. We construct
     spectral energy distributions, which we compare to different dust
     models, and calculate dust temperatures. We also derive opacity
     maps and analyze the density structure of the environment of
     HD\,97300.}
   {We find that HD\,97300 has no infrared excess at or below
     24$\,\mu$m, confirming its zero-age main-sequence nature. The
     morphology of the ring is very similar between 3.6$\,\mu$m and
     24$\,\mu$m. The emission at these wavelengths is dominated by
     either PAH features or PAH continuum. At longer wavelengths, only
     the northwestern part of the ring is visible. A fit to the
     100--500$\,\mu$m observations suggests that the emission is due
     to relatively warm ($\approx$26\,K) dust. The temperature
     gradually decreases with increasing distance from the ring. We
     find a general decrease in the density from north to south, and
     an approximate 10\% density increase in the northeastern part of
     the ring.}
   {Our results are consistent with the theory that the ring
     around HD\,97300 is essentially a bubble blown into the
     surrounding interstellar matter and heated by the star.}


   \maketitle


\section{Introduction}

At a distance of 160\,pc, Chamaeleon\,I is one of the closest and
richest star-forming regions, containing about 200 known low-mass
young stellar objects \citep{whittet1997,luhman2008}. There are only
two intermediate-mass stars in Chamaeleon\,I: HD\,97048 in the
southern part of the cloud and HD\,97300 in the northern part. While
HD\,97048 is a bona fide Herbig Ae/Be star, HD\,97300 seems to be
older, and is probably a zero-age main-sequence star \citep[e.g.][and
  references therein]{assendorp1990}. Excess emission with respect to
a B9-type stellar photosphere was detected at infrared wavelengths
towards HD\,97300, and was first attributed to an M-type companion or
a circumstellar dust shell
\citep{hyland1982,wesselius1984,the1986}. On the basis of maps of both
optical extinction and 130$\,\mu$m optical depth, \citet{jones1985}
found that HD\,97300 sits in the middle of a low-density region. They
interpreted this result in terms of a bubble cleared out by the (now
ended) stellar wind of HD\,97300.

\citet{prusti1994} later discovered that the 8--13$\,\mu$m emission is
extended, while \citet{ralf} were able to resolve the source with the
\emph{Infrared Space Observatory} at mid-infrared wavelengths, finding
that HD\,97300 is surrounded by an elliptical ring of size about
50$''\times$36$''$. The emission of the ring is dominated by emission
bands from polycyclic aromatic hydrocarbons (PAHs) at 6.2, 7.7, 8.7,
11.3, and 12.5$\,\mu$m (for a general review of PAHs, see
e.g.~\citealt{tielens2008}). \citet{ralf} modeled the ring with a
mixture of PAHs, very small grains, and large grains, and speculated
that the ring consists of interstellar matter entrained by either mass
loss or radiation pressure from the B9-type star. Using optical
polarization measurements, \citet{andersson2010} found evidence of a
local enhancement in the dust heating and radiatively driven grain
alignment.

In this paper, we present a comprehensive analysis of the surroundings
of HD\,97300 using infrared images covering a large wavelength range
from 3.6$\,\mu$m to 500$\,\mu$m. In Sect.~\ref{sec:observations}, we
briefly describe the reduction of the images taken with the
\emph{Herschel} Space Observatory \citep{pilbratt2010} and the
\emph{Spitzer} Space Telescope \citep{werner2004}. In
Sect.~\ref{sec:results}, we present images and brightness profiles
along different position angles. In Sect.~\ref{sec:discussion}, we
discuss the multiplicity of HD\,97300, and we analyze the spectral
energy distribution (SED) of the ring to constrain different dust
models. In Sect.~\ref{sec:summary}, we summarize our conclusions.

\section{Observations}
\label{sec:observations}

HD\,97300 was observed with \emph{Herschel} in January 2011 as part of
the ``Herschel Gould Belt survey'' \citep{andre2010}, using the PACS
\citep{poglitsch2010} and the SPIRE \citep{griffin2010}
instruments. Images at 70, 160, 250, 350, and 500$\,\mu$m images were
obtained simultaneously in SPIRE/PACS parallel mode. Additional
PACS-only 100 and 160$\,\mu$m images were also taken. Details of the
data reduction are described in papers by \citet{winston2012} and
\citet{vavrek2012}.

HD\,97300 was also observed with \emph{Spitzer} in July 2004. IRAC
images at 3.6, 4.5, 5.8, and 8.0$\,\mu$m were taken in High Dynamic
Range mode, alternating between exposure times of 10.4\,s and 0.4\,s
(aor 3651328). We downloaded the saturation-corrected basic calibrated
data (CBCD) produced by the pipeline version S18.18.0 at the Spitzer
Science Center. We used MOPEX \citep{makovoz2006} to create a mosaic
from the individual frames. Several pixels around HD\,97300 were
saturated in the long exposure mosaic. To ensure that we acquired data
of good signal-to-noise ratio for the faint extended emission, but
also correct fluxes for the bright point sources, we created a final
image by taking the long exposure image and replacing the pixels
affected by saturation with the corresponding pixels from the short
exposure image. Pixels unaffected by saturation had identical flux
values in the long and short exposure images within the measurement
uncertainties. MOPEX was also used to create mosaics from the MIPS
24$\,\mu$m and 70$\,\mu$m images (aor 3661312). At 24$\,\mu$m, we used
the basic calibrated data (BCD), while at 70$\,\mu$m, we used the
median-subtracted, time-filtered FBCD files. Exposure times for MIPS
were 3.67\,s at 24$\,\mu$m and 4.19\,s at 70$\,\mu$m, and on average
20 frames at 24$\,\mu$m and 10 frames at 70$\,\mu$m were used to
create the final mosaics. Saturation was not an issue at these
wavelengths.

\section{Results}
\label{sec:results}

\begin{figure*}
\centering
\includegraphics[width=17.6cm]{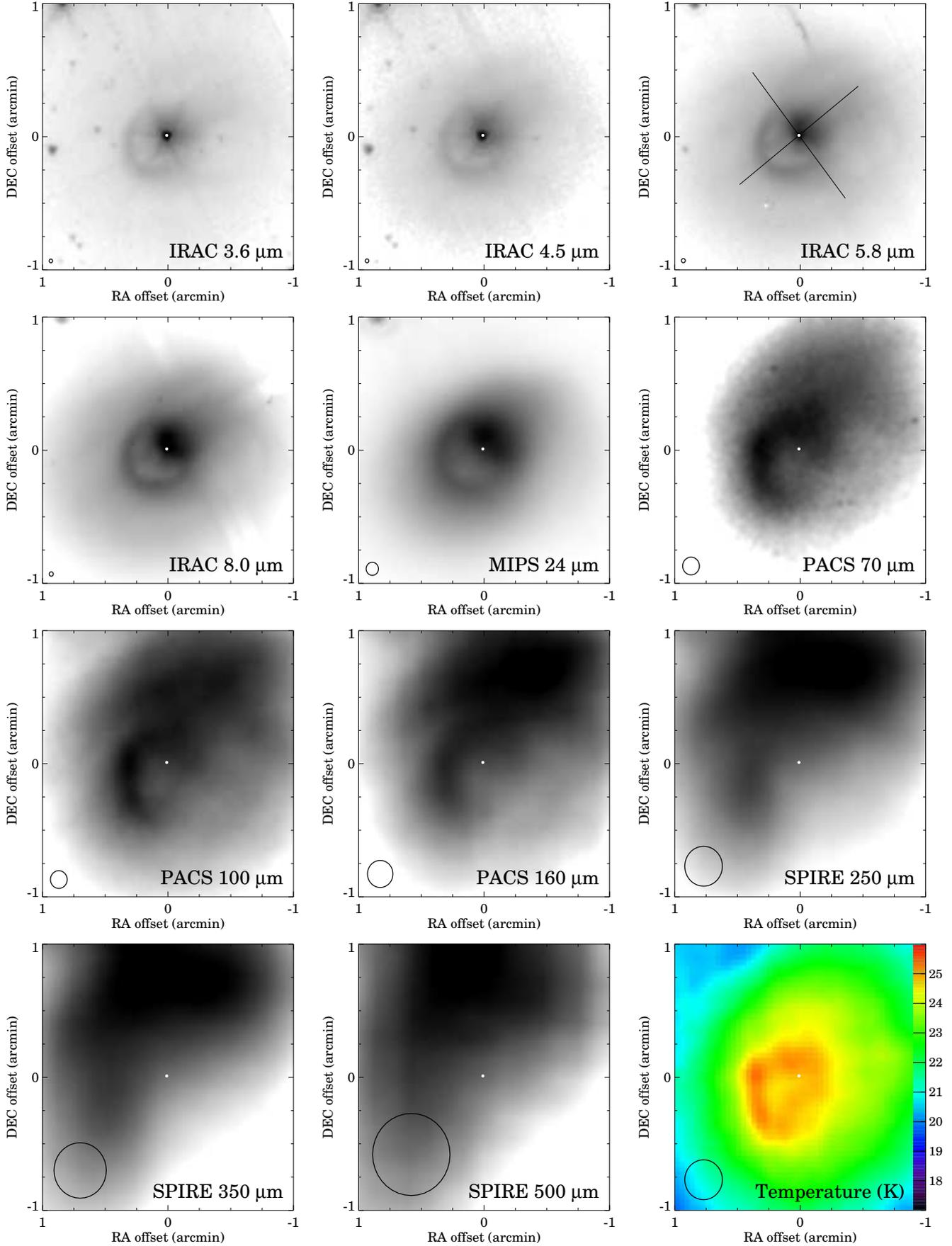}
   \caption{Images of HD 97300 at different wavelengths. The
     coordinates are relative to RA$_{2000}$=11$^{\rm h}$ 09$^{\rm m}$
     50$\fs$02 DEC$_{2000}$=$-$76$^{\circ}$ 36$'$ 47$\farcs$72. The
     scaling is logarithmic. Each panel is 2$'\times$2$'$, and north
     is up, east is left. The diameter of the circles in the bottom
     left corners corresponds to the FWHM of the PSF. The small white
     dot in each image indicates the stellar position. The lines in
     the upper right panel show the directions along which the cuts in
     Fig.~\ref{fig:cuts} were measured. The bottom right panel shows
     the dust temperature on a linear scale as measured by fitting a
     modified black body to the 100, 160, and 250$\,\mu$m images (see
     the text for details).}
      \label{fig:map}
\end{figure*}

Figure \ref{fig:map} presents 2$'\times$2$'$ images centered on
HD\,97300 at 11 different infrared wavelengths. We do not show the
MIPS 70$\,\mu$m image, because after convolving the higher spatial
resolution PACS 70$\,\mu$m image to match the resolution of the MIPS
70$\,\mu$m image (using the convolution kernels of
\citealt{gordon2008}), we found that they agree to within 20\%. At
160$\,\mu$m, we show merely the PACS-only mode image in
Fig.~\ref{fig:map}, because it is essentially identical to the
SPIRE/PACS parallel mode 160$\,\mu$m image. Our 3.6--24$\,\mu$m images
clearly show the same narrow elliptical ring around HD\,97300 imaged
by \citet{ralf} at 6.0--14.9$\,\mu$m. A point source coinciding with
HD\,97300 is seen between 3.6$\,\mu$m and 8.0$\,\mu$m, but a second
peak also appears at 8.0$\,\mu$m about 3--4$''$ north of
HD\,97300. This peak was previously detected by \citet{ralf} at
11.3$\,\mu$m and 14.9$\,\mu$m. The strength of HD\,97300 relative to
the ring gradually decreases with increasing wavelength and the star
is indeed invisible at 24$\,\mu$m and above. While the ring is
relatively symmetric (its northeastern part being similar to its
southwestern part) at shorter wavelengths, its northeastern part is
much more prominent at 70$\,\mu$m, an effect that is even more
pronounced at 100 and 160$\,\mu$m. A bright patch of emission to the
north of HD\,97300, probably originating in the Chamaeleon\,I cloud,
starts appearing at 100$\,\mu$m. At the SPIRE wavelengths, this patch
dominates the images, although emission from the northeastern part of
the ring remains visible.

\begin{figure}
\centering
\includegraphics[width=\columnwidth]{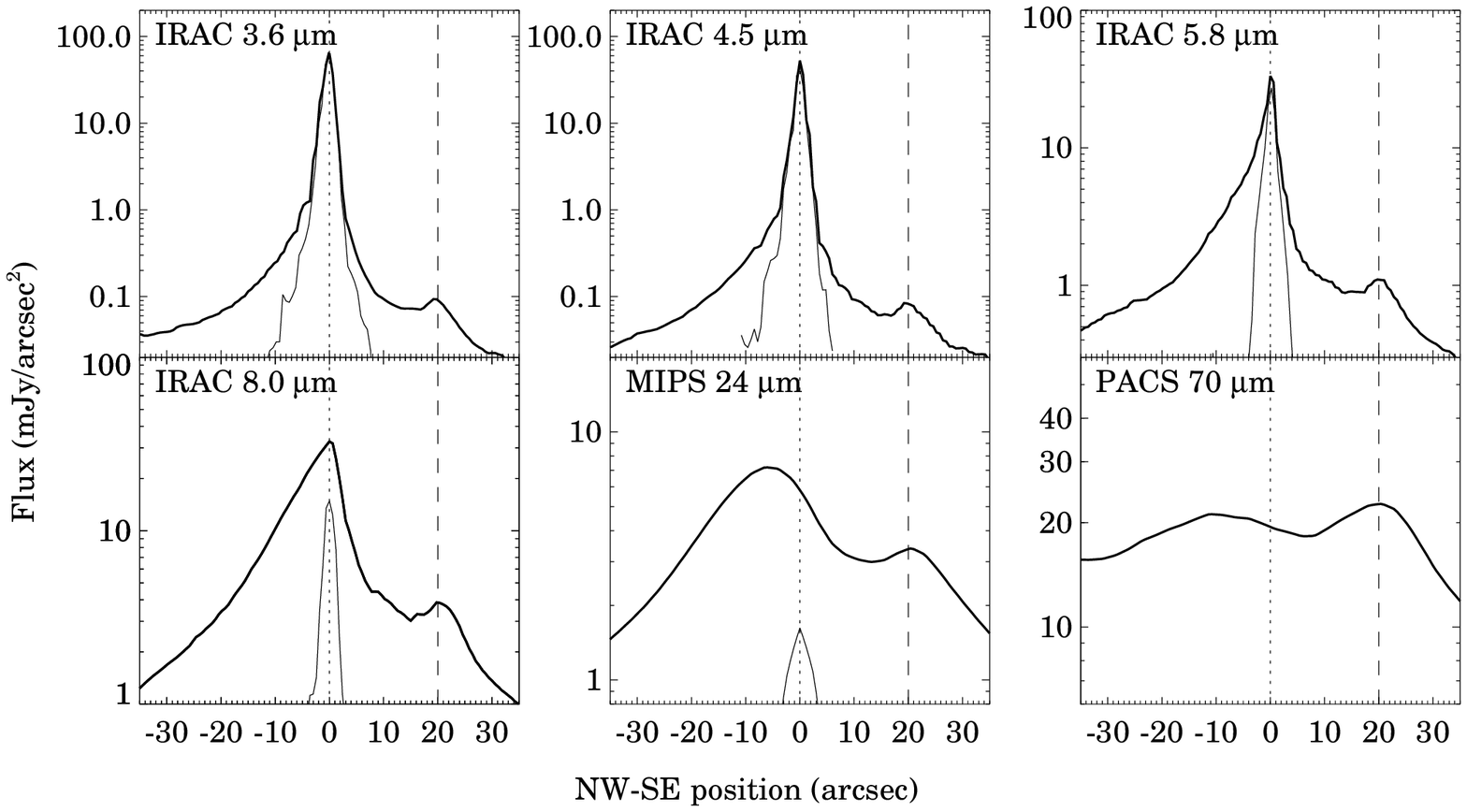}
\includegraphics[width=\columnwidth]{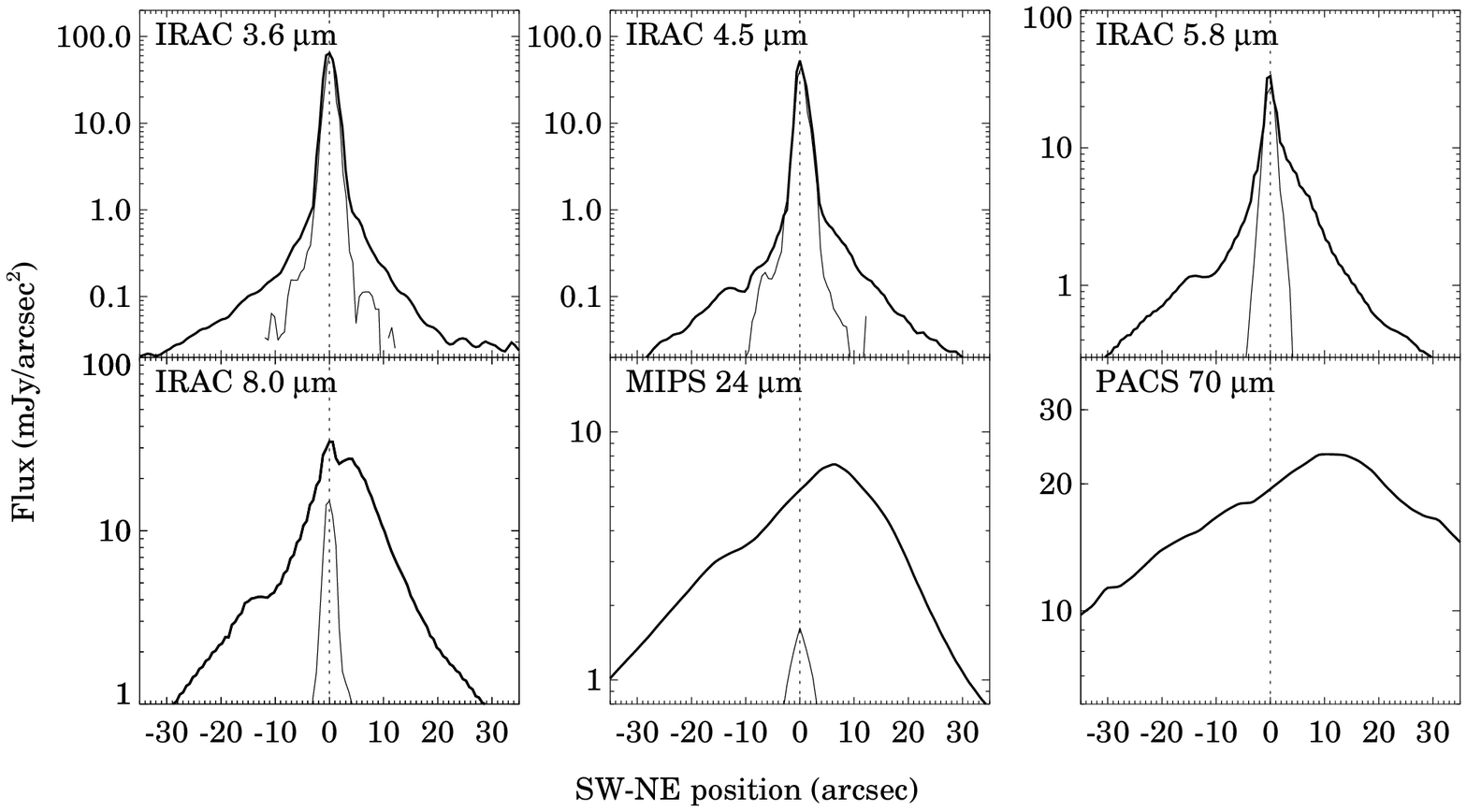}
   \caption{Logarithmic brightness profiles. The NW-SE cut was
     measured along a PA of 142$^{\circ}$, the SW-NE cut along a PA of
     52$^{\circ}$. These directions are marked with lines in the upper
     right panel of Fig.~\ref{fig:map}. The vertical dotted line
     indicates the stellar position, while the vertical dashed line
     indicates the ring. The thin lines show the PSF scaled to the
     expected stellar brightness (for more details, see the text).}
      \label{fig:cuts}
\end{figure}

To quantitatively analyze the images, in Fig.~\ref{fig:cuts} we
plotted the brightness profiles at different wavelengths along two
perpendicular lines across the star, one at a PA of 142$^{\circ}$
(NW--SE), the other one at 52$^{\circ}$ (SW--NE), as indicated by the
lines in the upper right panel of Fig.~\ref{fig:map}. For comparison,
we also plotted the observed point spread functions (PSFs), which were
constructed by taking the median of the normalized images of
four-to-six isolated stars in the field, then taking cuts along the
same PAs as mentioned before. Following \citet{ralf}, we scaled the
observed PSFs to ensure that at each wavelength they correspond to the
expected flux of a black body with a temperature of
10\,700\,K. Fig.~\ref{fig:cuts} shows that while at 3.6, 4.5, and
5.8$\,\mu$m the star dominates the observed flux (being at least an
order of magnitude brighter than the surrounding diffuse emission),
the star and the ring contribute almost comparably at 8.0$\,\mu$m, and
the stellar contribution is essentially negligible at 24$\,\mu$m and
above. Given that the observed profiles are consistent with HD\,97300
having a photospheric flux at 8$\,\mu$m and below, we can derive the
following stellar fluxes: F$_{3.6}\,{=}\,$0.41\,Jy,
F$_{4.5}\,{=}\,$0.28\,Jy, F$_{5.8}\,{=}\,$0.17\,Jy, and
F$_{8.0}\,{=}\,$0.09\,Jy. The uncertainty in these numbers is about
15\%. The error is dominated by the difficulty in determining the
precise background. From the non-detection of HD\,97400 at longer
wavelengths, we can infer upper limits to the stellar fluxes of
F$_{24}\,{<}\,$0.012\,Jy and F$_{70}\,{<}\,$0.06\,Jy (cf. the expected
photospheric fluxes of 0.011\,Jy at 24$\,\mu$m and 0.001\,Jy at
70$\,\mu$m). This result is consistent with the star having no
measurable infrared excess at or below 24$\,\mu$m. The SED of the star
is plotted with gray dots in the upper panel of Fig.~\ref{fig:ralf}.

Apart from the point source, extended emission is clearly seen in the
IRAC cuts plotted in Fig.~\ref{fig:cuts}. Starting from the northwest,
we see gradually increasing fluxes up to the point where the star is
situated. A steep decline follows, then there is another peak,
corresponding to the part of the ring farthest from the star. As
indicated by the dashed line in Fig.~\ref{fig:cuts}, this peak is
always located at a distance of 20$''$ from the star, irrespective of
the wavelength. Although the exact wavelengths are different, the
shape of our NW--SE cuts are in general consistent with the cuts
plotted in Fig.~4 of \citet{ralf} and we see no signs of the expansion
of the ring.

\section{Discussion}
\label{sec:discussion}

\subsection{Multiplicity of HD\,97300}

HD\,97300 is a subarcsecond binary with a separation of 0$\farcs$8, PA
of 327$^{\circ}$, and near-infrared magnitude differences of
$\Delta{}H$=3.33\,mag, $\Delta{}K$=3.05\,mag
\citep{ghez1997,lafreniere2008}. We do not resolve the binary system
at any infrared wavelength. Considering the near-infrared brightness
ratios and that the companion is probably a low-mass star, the
contribution of the companion to the photometry of HD\,97300 should be
at most 5--10\%. \citet{ralf} detected a second peak about 3--4$''$
north of HD\,97300 at 11.3$\,\mu$m and 14.9$\,\mu$m, although not at
shorter wavelengths, and speculated that it may be an embedded
companion. We also detected this peak at 8.0$\,\mu$m, but at our
spatial resolution, it is clearly resolved and not point-like (see
also the lower panel of Fig.~\ref{fig:cuts}, where the secondary peak
is visible on the IRAC 8.0$\,\mu$m cut at the position of about
4--5$''$). The lack of any source detected at this position in the
870$\,\mu$m continuum map of \citet{belloche2011} also argues against
the embedded companion scenario. Although we cannot completely exclude
the possibility that a core is associated with HD\,97300, the extended
nature of the peak and that it is only seen at 8.0, 11.3, and
14.9$\,\mu$m, strongly suggest that it is related to PAHs. It may be a
localized enhancement in PAHs, or an area where the excitation
conditions are more favorable.

\subsection{PAH emission}

\begin{figure}
\centering
\includegraphics[width=\columnwidth]{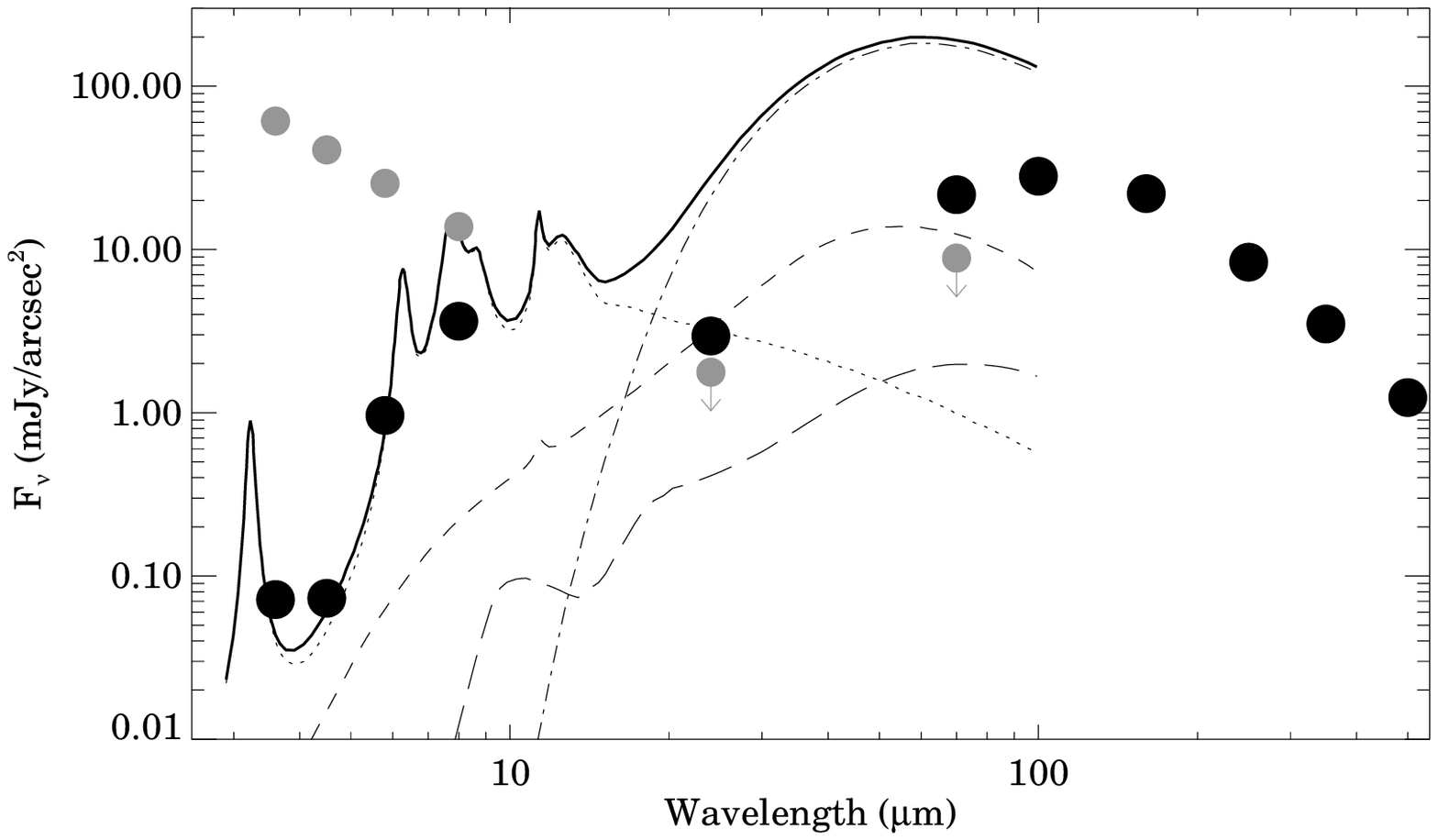}
\includegraphics[width=\columnwidth]{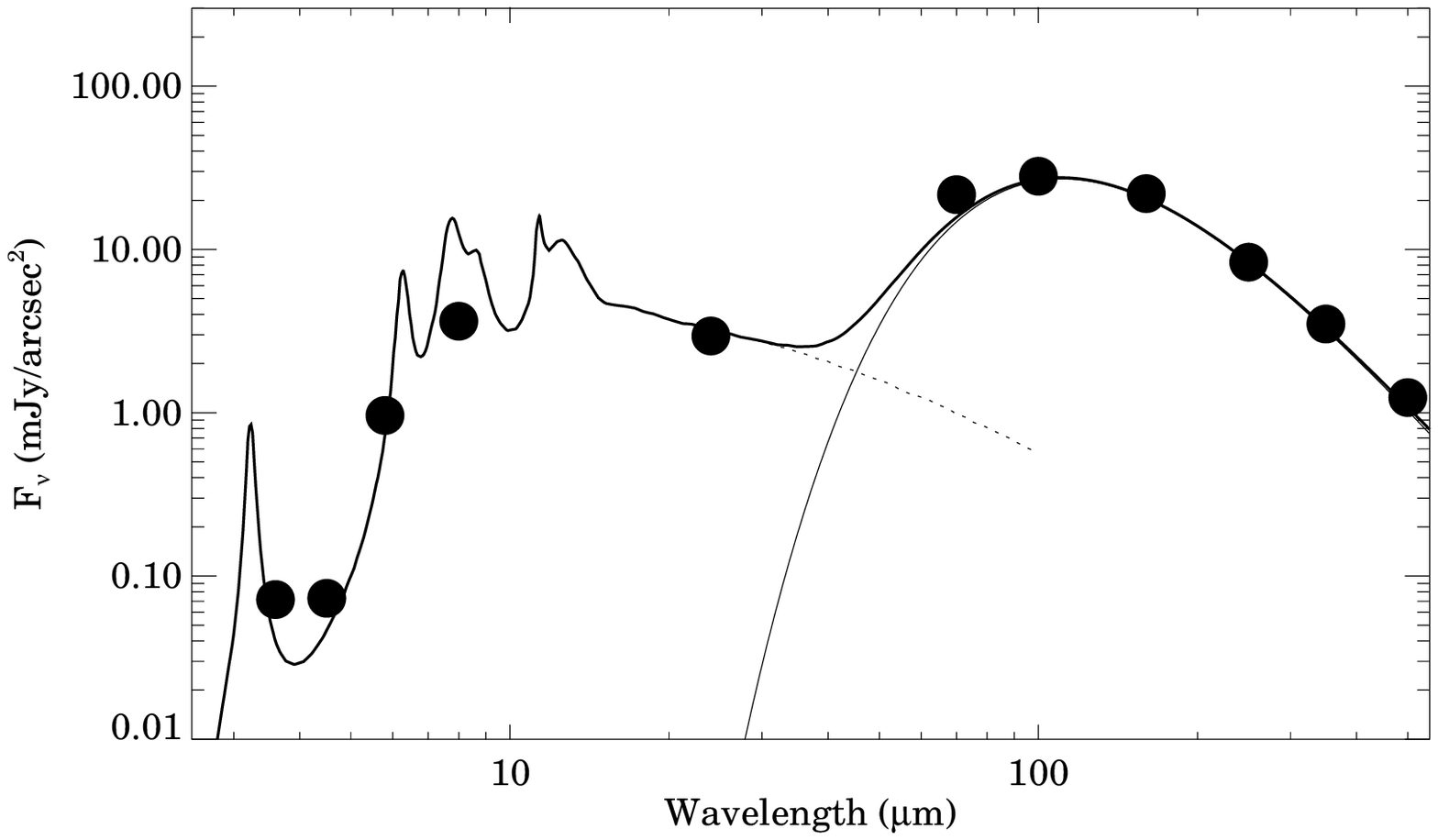}
   \caption{{\it Upper panel:} Dust model for the ring from
     \citet{ralf}. Dotted line: PAHs; dashed line: very small
     graphite; long dashed line: very small silicates; dash-dotted
     line: large grains. The small gray dots (from this work) mark the
     observational data or upper limits for the stellar photosphere,
     while the large black dots (also from this work) represent the
     ring emission as measured at a distance of 20$''$, PA of
     142$^{\circ}$ from the star (position 7 in \citealt{ralf}). Error
     bars are smaller than the symbol sizes, thus they are not
     plotted. {\it Lower panel:} the large black dots are the same
     observations as in the upper panel, but the overplotted model
     (thick solid line) is the sum of the PAHs from \citet{ralf}
     (dotted line) and a single-temperature modified black body (thin
     solid line) with $T$=26.0\,K and $\beta$=2.0.}
   \label{fig:ralf}
\end{figure}

Using spectrally resolved mid-infrared images, \citet{ralf} performed
a detailed analysis of the ring around HD\,97300. They fitted the
observed spectra with a mixture of PAHs, very small particles, and
large grains (Fig.~\ref{fig:ralf}, upper panel). They found that below
16$\,\mu$m, the emission is dominated by PAH features. Very small
graphite and silicate grains have at least a one order of magnitude
smaller contribution at all wavelengths. Above about 16$\,\mu$m, the
model is dominated by the thermal emission of large grains, although
continuum emission from PAHs is also included in their model. However,
in the absence of longer wavelength data, they were unable to check
the validity of the model at wavelengths longer than 16$\,\mu$m.

In the upper panel of Fig.~\ref{fig:ralf}, we reproduced Fig.~3 from
\citet{ralf}. We also overplotted the observed brightness of the ring
in our \emph{Spitzer} and \emph{Herschel} images at a distance of
20$''$, for the PA of 142$^{\circ}$ of the star, corresponding to the
position for which the dust model in Fig.~\ref{fig:ralf} was
calculated in \citet{ralf}. There is good agreement between the
observed brightness and the model for the four IRAC filters, but there
is a large discrepancy between the observations and the model at
longer wavelengths, where \citet{ralf} had no information about the
brightness of the ring. The IRAC 4.5$\,\mu$m filter mostly samples a
pseudo-continuum between the 3.3$\,\mu$m and 6.2$\,\mu$m PAH
features. The other three IRAC filters cover strong PAH features. In
agreement with the model, the extended emission in Fig.~\ref{fig:cuts}
indeed appears to be at its brightest in the 8$\,\mu$m image, which
includes the strong 7.7$\,\mu$m feature.

Judging from the dust model, the 24, 70, and 100$\,\mu$m images should
be the most well-suited to study the large grain population. On the
basis of the distance of the ring from the central star, \citet{ralf}
estimated a temperature of 50\,K and predicted that the emission would
peak at about 60$\,\mu$m. Our observations, however, suggest that the
peak is instead somewhere between 70$\,\mu$m and 160$\,\mu$m,
indicating that there is no warm (50\,K) dust in the system. The
observed 24$\,\mu$m brightness can still be fully explained by
PAHs. The morphology of the ring between 3.6$\,\mu$m and 24$\,\mu$m
remains almost unchanged, supporting the idea that the emission in
this wavelength range is coming almost exclusively from PAH molecules.

\subsection{Warm dust}

At 70$\,\mu$m and above, PAH emission can no longer explain the
observed SED of the ring. The upper panel of Fig.~\ref{fig:ralf} shows
that we need colder ($<$50\,K) dust to fit the observations. If we
take into account data points from 100$\,\mu$m to 500$\,\mu$m, the SED
can be well-fitted with a modified black body (Fig.~\ref{fig:ralf},
lower panel). The best-fit model has a temperature of $T$=26.0\,K for
a fixed power-law index of the dust opacity coefficient $\beta$=2.0
(assuming a dust opacity law of
$\kappa\,{\propto}\,\lambda^{-\beta}$). For the diffuse interstellar
medium, $\beta\,{\approx}\,$1.8 \citep{draine2006}, while grain growth
results in $\beta\,{\lesssim}\,$1 \citep{beckwith1991}. That a black
body with $\beta$=2 fits the observed data points very well suggests
that the ring around HD\,97300 contains interstellar medium-like
dust. Interestingly, the ratio of the total to selective extinction,
$R_V=A_V/E(B-V)$ for HD\,97300 is around 4.9--5.5
\citep{grasdalen1975,steenman1989b}. For the diffuse interstellar
medium, the typical value of $R_V$ is 3.1, while grain growth leads to
larger $R_V$ values \citep{steenman1989,cardelli1989}. This is not
necessarily a contradiction for the following reason. The value of
$R_V$ is very sensitive to small variations in the grain size. A
change of the maximal grain size from 0.22$\,\mu$m to 1$\,\mu$m
results in an increase of $R_V$ from 3.1 to 5.9 \citep[see Fig.~3
  in][]{steenman1989}. The value of $\beta$ derived from the slope of
the submillimeter SED, on the other hand, only changes significantly
if the grains grow to several tens of microns \citep[see Fig.~3
  in][]{draine2006}. Thus, some grain growth has occurred in the
vicinity of HD\,97300, but the grains remain no larger than a
micrometer size. We note that another solution to the difficulty in
reconciling the observed $R_V$ and $\beta$ may be that they trace
different parcels of dust. While optical and near-infrared data
(yielding $R_V$) probe only line-of-sight material in front of the
star, far-infrared data (yielding $\beta$) might contain contributions
from the dust in the underlying cloud.

Modified black-body fits could be obtained not only for the ring, but
for any point in our data map of sufficiently high signal-to-noise
ratio. We used the convolution kernels of \citet{gordon2008} to smooth
the 100$\,\mu$m and 160$\,\mu$m images to the resolution of the
250$\,\mu$m image, and used these three images to fit modified
black-body curves for each pixel with $\beta$ fixed at 2.0. The
resulting temperature map is plotted in the bottom right panel of
Fig.~\ref{fig:map}. The ring around HD\,97300 clearly stands out in
the temperature map, with temperatures of around 25--26\,K. The
temperature gradually decreases with increasing distance from the ring
to about 20--23\,K. The morphology of the temperature map suggests
that the dust in the ring around HD\,97300 is heated by the central
star, but not so much as the 50\,K expected by
\citet{ralf}. Interestingly, the dust temperature in the vicinity of
HD\,97300 is still higher than the typical temperatures in the
Chamaeleon complex: \citet{toth2000} found that the intercloud
material has a typical dust color temperature of $\approx$16.3\,K,
while the obscured clouds are $\le$14.5\,K. \citet{ikeda2012} found
two dust components in Chamaeleon\,I, a cold one with a temperature of
11.7\,K, and a warm one with a temperature of 22.1\,K. Thus, the
immediate surroundings of HD\,97300 are slightly warmer even than this
warm component.

\subsection{Density structure}

Our results suggest that the long wavelength emission of the ring
around HD\,97300 originates in relatively warm ($\approx$26\,K)
dust. Interestingly, practically the full ring is visible in the
temperature map, while only the northeastern arc is clearly visible in
70--250$\,\mu$m emission. This indicates that the dust density is
probably higher in the northeastern part of the ring than in the
southwestern part. To test this hypothesis, we derived opacity maps
($\tau_{100}$ and $\tau_{160}$) using the temperature map and the
100$\,\mu$m and 160$\,\mu$m surface brightness maps plotted in
Fig.~\ref{fig:map}. In terms of the large-scale structure, our opacity
maps are remarkably similar to the $\tau_{130}$ map of
\citet{jones1985}. The opacity is lower around HD\,97300 and towards
the south ($\tau_{100}\,{\approx}\,$0.8$\times$10$^{-3}$), and higher
farther from the star and to the north (up to
$\tau_{100}\,{\approx}\,$1.8$\times$10$^{-3}$ in the studied
2$'\,{\times}\,$2$'$ area). An important difference with respect to
the opacity map of \citet{jones1985} is that the high spatial
resolution of our \emph{Herschel} maps also enables us to study the
density structure of the ring. Only the northeastern part of the ring
is visible in our opacity maps, indicating that it is indeed denser
than the southwestern part. The opacity (and thus density) contrast
between the northeastern part of the ring and its surroundings is
about 10\%.

By averaging the opacity in a region within 33$''$ of the star, we get
$\tau_{100}$=1$\times$10$^{-3}$. Using $A_V/\tau_{100}$=1000\,mag as a
typical value for the diffuse interstellar medium \citep{kiss2006}, we
can convert the far-infrared opacity to the optical extinction of
$A_V$=1\,mag. Taking the relation between the optical extinction and
the hydrogen column density from \citet{guver2009}, we obtain
$N_H$=2.2$\times$10$^{21}$\,cm$^{-2}$ (or
3.7$\times$10$^{-3}$\,g\,cm$^{-2}$). This gives a total mass of
0.008\,M$_{\odot}$ for this region. The mass in the ring is probably
less than 10\% of this value, that is about an order of magnitude less
than the mass determined by \citet{ralf} from their PAH observations
(0.07\,M$_{\odot}$ for the whole 33$''$-radius area, and
0.03\,M$_{\odot}$ for the ring). Since these are order of magnitude
calculations rather than precise modeling, these values are not
inconsistent, but indicate the uncertainty in the estimates.

The temperature smoothly decreases farther from the ring, with no
significant difference between the northern and southern sides. Thus,
the bright patch of emission seen to the north of HD\,97300 in the
long wavelength maps in Fig.~\ref{fig:map} probably also indicates a
density enhancement to the north. The 870$\,\mu$m continuum map of
\citet{belloche2011}, as well as our opacity maps also suggest that
there is a general increase in column density towards the north in the
area of HD\,97300. These findings support the theory that the ring
around HD\,97300 is essentially a bubble blown into the interstellar
matter by the central star, either due to its stellar wind or its
radiation pressure. The density ($\rho$) gradient of the interstellar
dust in the vicinity of HD\,97300 naturally explains the asymmetric
and nonconcentric appearance of the ring, since the radius of a
stellar wind bubble is proportional to $\rho^{-1.5}$
\citep{weaver1977}.

\section{Summary}
\label{sec:summary}

We have presented infrared images of HD\,97300 and its surroundings at
11 different wavelengths between 3.6$\,\mu$m and 500$\,\mu$m. By
analyzing the brightness profiles across the star along different
position angles, we have found that the star has no infrared excess at
or below 24$\,\mu$m, confirming its zero-age main-sequence nature. We
constructed the SED of the ring based on our 11 data points and
compared it to the dust model of \citet{ralf}. On the basis of the SED
and the morphology of the emission at different wavelengths, we have
concluded that our data set can be explained in terms of a
two-component dust model. The emission up to 24$\,\mu$m is consistent
with that from PAH molecules, while at longer wavelengths, we observe
thermal emission from relatively warm ($\approx$26\,K) dust. The
emission above 70$\,\mu$m suggests that there is an inhomogeneous
distribution of the interstellar matter, and we conclude that the ring
around HD\,97300 is probably a bubble blown into the interstellar
matter and heated by the star.


\begin{acknowledgements}

PACS has been developed by a consortium of institutes led by MPE
(Germany) and including: UVIE (Austria); KUL, CSL, IMEC (Belgium);
CEA, OAMP (France); MPIA (Germany); IFSI, OAP/AOT, OAA/CAISMI, LENS,
SISSA (Italy); IAC (Spain). This development has been supported by the
funding agencies BMVIT (Austria), ESA-PRODEX (Belgium), CEA/CNES
(France), DLR (Germany), ASI (Italy), and CICT/MCT (Spain)."

SPIRE has been developed by a consortium of institutes led by Cardiff
Univ. (UK) and including: Univ. Lethbridge (Canada); NAOC (China);
CEA, LAM (France); IFSI, Univ. Padua (Italy); IAC (Spain); Stockholm
Observatory (Sweden); Imperial College London, RAL, UCL-MSSL, UKATC,
Univ. Sussex (UK); and Caltech, JPL, NHSC, Univ. Colorado (USA). This
development has been supported by national funding agencies: CSA
(Canada); NAOC (China); CEA, CNES, CNRS (France); ASI (Italy); MCINN
(Spain); SNSB (Sweden); STFC (UK); and NASA (USA).

PR and NLJC acknowledge support from the Belgian Federal Science
Policy Office via the PRODEX Programme of ESA.

\end{acknowledgements}

\bibliographystyle{aa}
\bibliography{paper}{}

\end{document}